\documentclass[a4paper,11pt]{article}
\pdfoutput=1 

\usepackage{jheppub} 

\usepackage[T1]{fontenc} 
\usepackage[utf8]{inputenc}
\usepackage[english]{babel}

\usepackage{color, colortbl}
\DeclareMathOperator{\arctanh}{arctanh}

\usepackage{./gnuplot-lua-tikz}
\usepackage{subfig}

\title{\boldmath Sgluons in the same-sign lepton searches}

\author[]{Wojciech Kotlarski
}

\affiliation[]{Faculty of Physics, University of Warsaw,\\ Pasteura 5, 02093 Warsaw, Poland}
\affiliation[]{
Institute of Nuclear and Particle Physics, TU Dresden\\
 01069 Dresden, Germany}

\emailAdd{wojciech.kotlarski@fuw.edu.pl}

\abstract{
In this work I present the interpretation of the ATLAS search of same-sign lepton production in association with $b$-jets in the context of the 4-top quark signal from sgluon decays.
I show that using just 3.2/fb data sample from Run 2 the exclusion limit is already competitive with the Run 1 one.
Public data allow to exclude sgluons with masses up to 0.95 TeV. 
Prospects for the total Run 2 integrated luminosity of 100/fb are briefly discussed.
}

\begin{document} 
\maketitle
\flushbottom

\section{Introduction}

With the Large Hadron Collider (LHC) delivering data at an unprecedented energy of 13 TeV a lot of work has been devoted to their interpretation in the context of BSM physics.
For the time being, the main focus is on the Minimal Supersymmetric Standard Model (MSSM) or the so-called exotics.
This of course leaves a lot of interesting models out.
From the viewpoint of supersymmetry, this is a serious limitation.
Recent years brought a lot of attention to the extended SUSY models, from simple extensions, like the NMSSM, to models with an extended QCD sector like, for example, various models with Dirac gluinos.
Studies proved that MSSM bounds are in many cases not applicable to these models~\cite{Heikinheimo:2011fk}.
On the other hand, 13 TeV data might be already more constraining than the 7 and 8 TeV one, even though the collected integrated luminosity is smaller.
This then raises an important question about the validity of such models in light of new data.

Especially interesting are the multi top-quark processes which, while characterized by a high mass scale, enjoy a big boost when going from 7 or 8 to 13 TeV.
The 4-top quark final state was already searched for by ATLAS~\cite{Aad:2016tuk,ATLAS-CONF-2016-032,ATLAS-CONF-2016-020,ATLAS-CONF-2016-013} and CMS~\cite{Khachatryan:2016kod,Khachatryan:2016uwr,Khachatryan:2016kdk} at Run 2. 
In the MSSM this kind of final state may appear as decay products of 3rd generation stops (produced either directly or as decay products of intermediate gluinos).
In general SUSY models the resonance structure might be quite different, though.
One might for example expect a two body decay of a new color resonance directly to a $t\bar{t}$ pair.
This is a general feature of models containing color octet (EW-singlet) scalars, commonly dubbed sgluons.
Their LHC phenomenology was previously investigated in the context of \textit{R}-symmetric/$\mathcal{N}=2$/Dirac gaugino SUSY models, hyper-pions in vector-like confinement gauge theories and universal extra dimensions~\cite{Choi:2008ub,Plehn:2008ae,Calvet:2012rk,Kilic:2008ub,Schumann:2011ji,Chen:2014haa,Beck:2015cga,Degrande:2014sta,Dobrescu:2007yp,Kilic:2010et,Burdman:2006gy,Benakli:2016ybe}.

The Minimal \textit{R}-Symmetric Supersymmetric Standard Model (MRSSM)~\cite{Kribs:2007ac} is a particularly well motivated BSM model~\cite{Randall:1992cq,Bertuzzo:2014bwa,Ellis:2016gxa,Braathen:2016mmb,Diessner:2014ksa,Diessner:2015yna,Diessner:2015iln}. 
Recent analyses \cite{Diessner:2014ksa,Diessner:2015yna} showed at full one- and leading two-loop levels that the 125 GeV Higgs boson can be consistently obtained in agreement with precision EW observables and flavor constraints. 
Moreover, interesting scenarios have been identified \cite{Diessner:2015iln} which provide a viable candidate for dark matter. 

Within the framework of the MRSSM~\cite{Kribs:2007ac} sgluons are expected to decay, depending on theirs mass, mainly into gluons or top quarks.
This kind of signatures, in both channels, were searched for by the experimental collaborations in 7 and 8 TeV data.
ATLAS excludes at 95\% CL pair produced, complex sgluons decaying (with branching ration 1) to gluon pair in mass range from 100 to 287 GeV \cite{ATLAS:2012ds}.
For $t\bar{t}$ decay mode, sgluons are excluded at 95\% CL up to 1.06 TeV  \cite{Aad:2015kqa}.
It should be noted though, that these exclusions are based on the simplified model with a complex sgluon from Ref.~\cite{GoncalvesNetto:2012nt} while in the MRSSM the cross section is roughly 2 times smaller.\footnote{ATLAS analysis also does not specify the form of the sgluon - top quark coupling.
}
At the time of writing there are no 13 TeV analyses addressing directly sgluon pair production.
Therefore, all mentioned exclusions come from Run 1.
This makes any projections for the target Run 2 integrated luminosity very difficult.  
To fill this gap, this work recasts current ATLAS limits from search of SUSY in the 4-top quark final state in Ref.~\cite{Aad:2016tuk} to sgluon pair production.

The paper is structured as follows.
The next section describes and motives the effective sgluon model used in this work.
Section~3 presents NLO cross sections for the sgluon pair production.
In section 4 the setup for the Monte Carlo simulation is described.
Section 5 then describes the parametrization of the detector response and the encoded ATLAS analysis.
The reproduced analysis is then validated on the associated production of top quark pair and a gauge boson, comparing predicted numbers of background events with the ones quoted by the ATLAS work.
The analysis is then applied to the signal events.
The work finishes with the derivation of the limit on the sgluon mass and prospects for this limit for the predicted $\gtrsim 100~\text{fb}^{-1}$ data sample of Run 2.

\section{Description of the model}

I work in the framework of a simplified model inspired by the MRSSM scenario in which all the superpartners but the \textit{CP}-odd sgluon are heavy.
The Standard Model (SM) gets extended by a real color-octet (EW-singlet) scalar $O$.
It couples exclusively to gluons and top quarks as given by the Lagrangian
\begin{equation}
\mathcal{L} = \mathcal{L}^{\text{SM}} + \frac{1}{2} D_\mu O^a D^\mu O^a - \frac{1}{2} m^2_{O} O^2 - \imath c \bar{t} \gamma^5 T^a t O^a,
\label{eq:lagrangian}
\end{equation}
where $D_\mu$ is the $SU(3)_C$ covariant derivative and sum over the color index $a$ is understood.
This is motivated by the MRSSM particle spectrum in which a complex sgluon field gets split into \textit{CP}-even and odd components through a $D$-term SUSY breaking contribution~\cite{Diessner:2015yna}.\footnote{I neglect possible (anti-)holomorphic soft-breaking sgluons mass terms.}
The masses of the components are then $m_{O_S}^2 = m_O^2 + 4 (M_O^D)^2$ for the scalar and $m_{O_A}^2 = m_O^2$ for the pseudoscalar, where $m_O$ and $M_O^D$ are sgluon and Dirac gluino soft masses.
Since physical gluino mass, which at the tree-level is exclusively controlled by the $M_O^D$, must be $\gtrsim 1$ TeV this implies that either pseudoscalar sgluon is very light and scalar one is in a TeV range or, if pseudoscalars mass is around 1 TeV, a scalar one will be in the multi-TeV range.
Here I focus on the latter scenario extending the SM with a pseudoscalar sgluon which for simplicity I denote just by $O$ (without the $A$ subscript).

\begin{figure}
  \centering
  \includegraphics[width=0.6\textwidth]{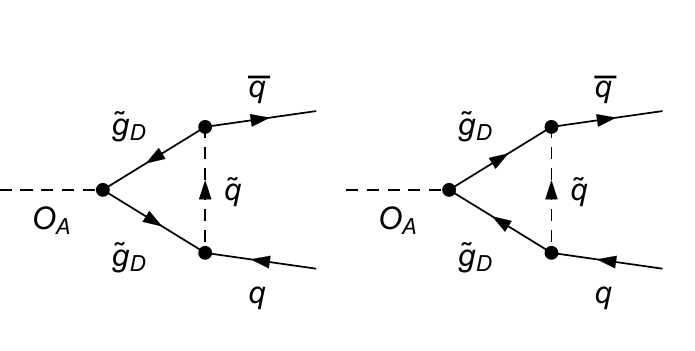}
  \caption{Lowest order diagrams generating (effective) coupling of pseudoscalar sgluon $O_A$ to quarks.}
  \label{fig:Op2qqbar}
\end{figure}

Since in the MRSSM sgluon carries an R-charge 0, once produced it can decay to SM particles.
The lowest order coupling to quarks is loop-induced as show in Fig.~\ref{fig:Op2qqbar}.
The coupling to gluons vanishes for pseudoscalar sgluons while the coupling to quarks is proportional to quark mass due to chirality.
Pseudoscalar sgluons with mass $m_{O_A} \gtrsim 2 \, m_t$ and smaller than other color-charged SUSY particles will therefore decay almost exclusively to top quarks with the coupling of the form written in Eq.~\ref{eq:lagrangian}.
This also motivates why I do not consider a single sgluon production through (loop-induced) coupling to partons.
This occurs mainly through coupling of gluons to the \textit{CP}-even one, which is significantly heavier than the \textit{CP}-odd one and whose production is additionally suppressed by a small value of the loop-induced coupling.

It should be noted though that the effective model described by the Lagrangian from Eq.~\ref{eq:lagrangian} is quite generic and can come from a multitude of complete, high scale theories.
Different models would then by characterized by a different chiral structure of the coupling $c$, though.

\section{NLO QCD corrections to sgluon pair production}

\begin{figure}
  \centering
  \includegraphics[width=0.76\textwidth]{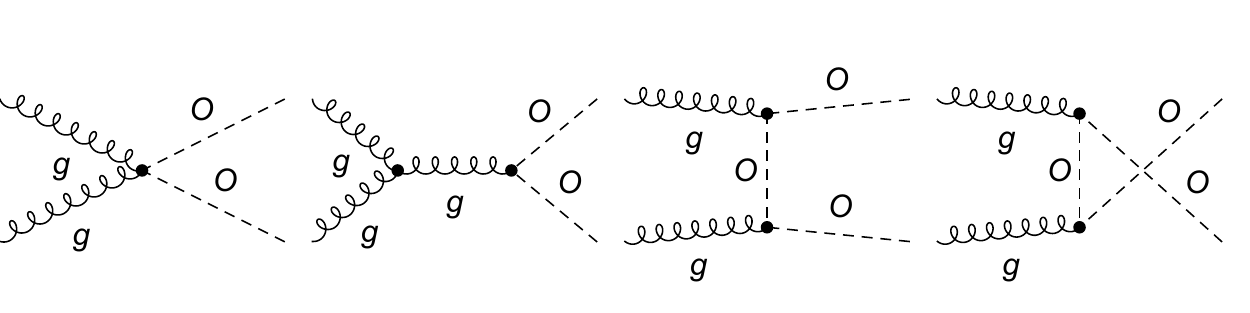}
  \includegraphics[width=0.19\textwidth]{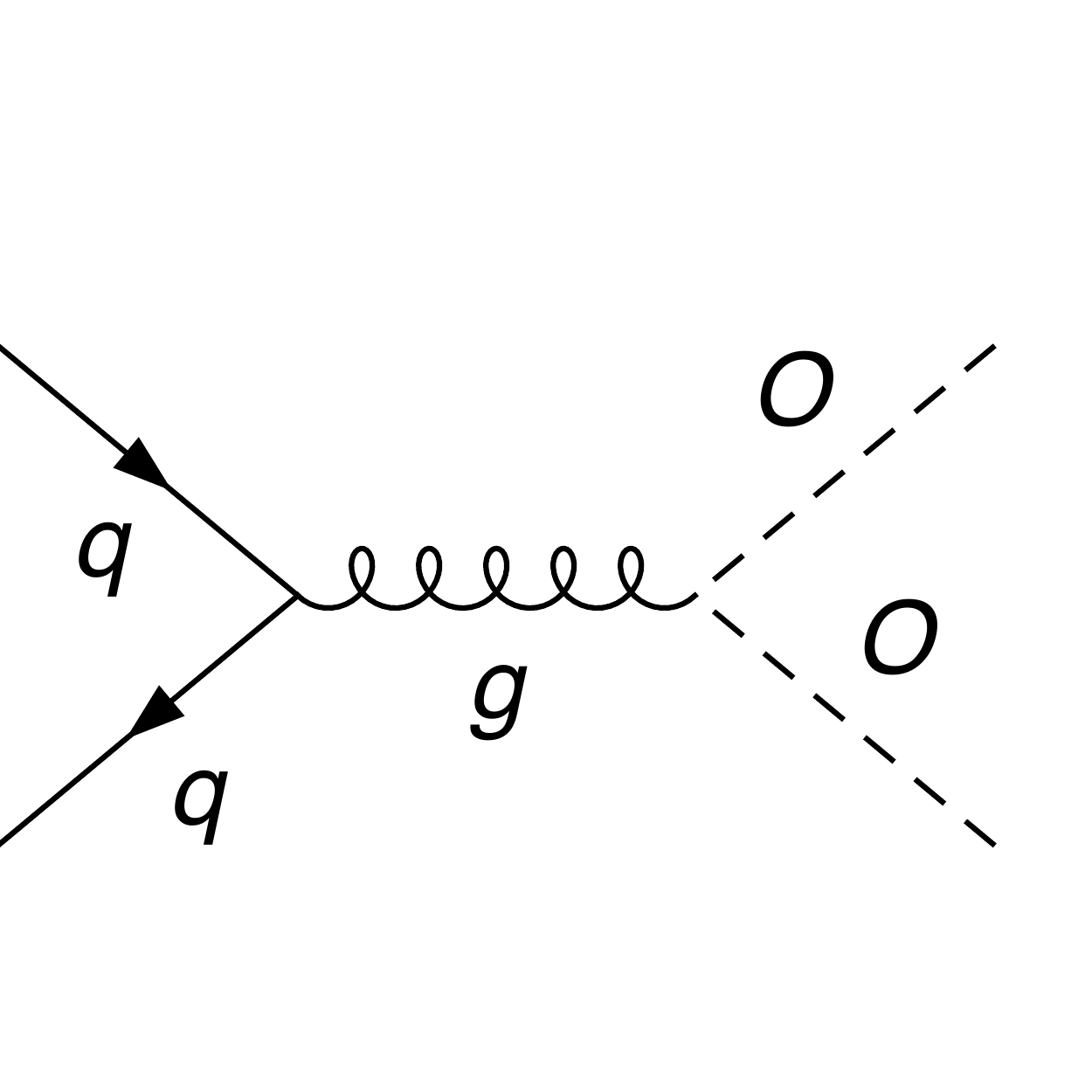}
  \caption{Feynman diagrams for the sgluon pair production at the LO. \label{fig:sgluon_feynman_diagrams_@LO}}
\end{figure}

For the Lagrangian of Eq.~\ref{eq:lagrangian} sgluons are produced at the LO through Feynman diagrams in Fig.~\ref{fig:sgluon_feynman_diagrams_@LO}.
The corresponding partonic cross sections are\footnote{With an additional factor of 1/2 compared to cross sections for a complex sgluon pair production considered in Ref.~\cite{Choi:2008ub}.}:
\begin{align}
\label{eq:lo_crosssection1}
  \hat{\sigma}_{q \bar{q}}^B & = \frac{2\pi\alpha_s^2}{9\hat{s}} \beta^3 ,\\
\label{eq:lo_crosssection2}
  \hat{\sigma}_{g g}^B & = \frac{3 \pi \alpha_s^2}{32 \hat{s}} \left ( 27 \beta - 17 \beta^3 +6 (-3 + 2\beta^2 + \beta^4 )\arctanh \beta \right  ),
\end{align}
where $\hat s \equiv (p_q + p_{\bar{q}})^2$ or $(p_g + p_{g'})^2$ and $\beta$ is sgluon's velocity in the center of mass system of colliding partons.

The first calculation of higher order corrections to the sgluon pair production was done in Ref.~\cite{GoncalvesNetto:2012nt} for a simplified model with a complex sgluon.
Since Ref.~\cite{Degrande:2014sta} a general procedure for obtaining NLO capable UFO~\cite{Degrande:2011ua} models for \texttt{MadGraph5\_aMC@NLO}~\cite{Alwall:2014hca} using conjunction of \texttt{FeynRules}~\cite{Alloul:2013bka}, \texttt{NLOCT}~\cite{Degrande:2014vpa}, \texttt{FeynArts}~\cite{Hahn:2000kx} and \texttt{FormCalc}~\cite{Hahn:1998yk} became available.
In Ref.~\cite{Degrande:2014sta} this procedure was applied to, among others, obtaining an NLO model for a real sgluon field.
Since the original model, available under \url{https://feynrules.irmp.ucl.ac.be/wiki/NLOModels}, does not work for the complex coupling $\imath c$ as in Eq.~\ref{eq:lagrangian}, a new model (this time in 5-flavor scheme) was generated and used for the analysis below.\footnote{The NLO UFO model used for this analysis can be found in supplementary materials with the \texttt{arXiv} version of this work. }

\begin{figure}
  \centering
  \input{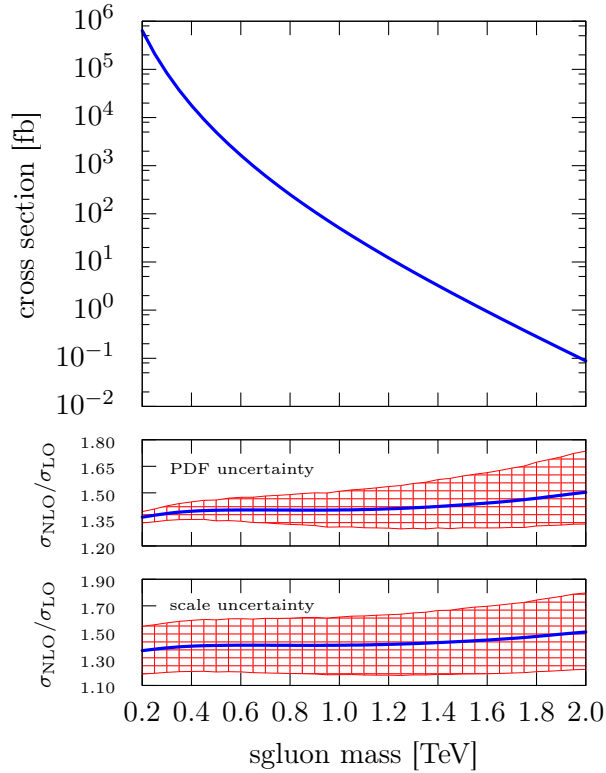}  
  \caption{NLO cross section for the sgluon pair production as a function of their mass.
  Middle subfigure shows the $K$-factor (blue line) together with the uncertainty band coming from the PDFs.
  Lower one does the same for the uncertainty coming from the scale variation.
  \label{fig:sgluon_xsection_NLO}
  }
\end{figure}

Table~\ref{tab:sgluon_xsection_NLO} lists values of cross sections obtained with this model for 5 selected sgluon massed: 1, 1.25, 1.5, 1.75 and 2 TeV, for 13 and 14 TeV LHC.
Numbers were obtained using the \texttt{MMTH2014} baseline (5-flavor) NLO fit (\texttt{MMTH2014nlo68cl}) \cite{Harland-Lang:2014zoa} interfaced through \texttt{LHAPDF6}~\cite{Buckley:2014ana}.
The $K$-factors listed in the table are defined as $K \equiv \sigma_{\text{NLO}}/\sigma_{\text{LO}}$ and refer to the LO calculation with \texttt{MMTH2014} baseline LO fit with $\alpha_s(m_Z) = 0.135$ and up to 5 active flavors (\texttt{MMTH2014lo68cl}).
For the sgluon with mass of 1 TeV one expects more than 100 events already with the publicly available data sample of 3.2 fb$^{-1}$.
Figure \ref{fig:sgluon_xsection_NLO} shows the plot of the cross section as a function of the sgluon mass together with uncertainty bands for the $K$-factor coming from the PDFs (middle subplot) and variation of renormalization/factorization scales by a factor of 2 (bottom subplot).
The central values of renormalization and factorization scales are set equal to the sgluons mass while $m_t = 173$ GeV.

Results of an automated  \texttt{MadGraph5\_aMC@NLO} calculation were cross-checked with an independent computation based on \texttt{FeynArts}, \texttt{FormCalc} and the two-cut phase space slicing (TCPS) method \cite{Harris:2001sx}.
The details of this computation will become available in a separate publication~\cite{my_phd}.
For the full description of the TCPS method with its application to the calculation of (S)QCD corrections to squark pair production in the MRSSM I also refer to the forthcoming publication \cite{our_paper}.

\begin{table}
  \centering
  \begin{tabular}{c|cc|cc}
    sgluon mass [TeV] & cross section at 13 TeV [fb] & $K$ & cross section at 14 TeV [fb] & $K$\\
    \hline
    1  & $50.79^{+15.3\%+7.7\%}_{-15.7\%-6.7\%}$ & 1.40 & $71.41^{+14.1\%+7.2\%}_{-15\%-6.3\%}$ & 1.37\\
    1.25 & $8.656^{+16.3\%+9.5\%}_{-16.5\%-7.9\%}$ & 1.38 & $12.91^{+14.9\%+8.8\%}_{-15.7\%-7.4\%}$ & 1.41 \\
    1.5 & $1.726^{+17.3\%+11.3\%}_{-17.2\%-9.1\%}$ & 1.40 & $2.752^{+15.8\%+10.5\%}_{-16.3\%-8.5\%}$ & 1.39\\
    1.75 & $0.3797^{+18.4\%+13.3\%}_{-17.9\%-10.5\%}$ & 1.46 & $0.6482^{+16.7\%+12.3\%}_{-17\%-9.7\%}$ & 1.41 \\
    2 & $0.08832^{+19.7\%+15.5\%}_{-18.8\%-11.9\%}$ & 1.47 & $0.1635^{+17.8\%+14.2\%}_{-16.5\%-11\%}$ & 1.45 \\
  \end{tabular}
  \caption{
    Cross sections for the sgluon pair production for 13 and 14 TeV LHC as a function of the sgluon mass (see main text for more details).
    First error comes from the scale variation, second is the PDF uncertainty (evaluated over PDF eigenvectors using hessian method).
    Relative statistical errors are below $10^{-3}$ and not shown here.
    Column $K$ gives global $K$-factors.
    \label{tab:sgluon_xsection_NLO}
    }
\end{table}

\section{Monte Carlo simulation setup}
I now proceed to the description of the methods used in the simulation of signal and background processes.
Due to technical reasons, samples for signal and background were generated using two different methods outlined in the next two subsections.
Both for signal and background simulation, following values of SM gauge-boson masses were used:
$m_W = 80.385$~GeV, $m_Z = 91.1876$~GeV. 
Top quark mass was set to 173.21 GeV while other quarks were assumed massless in the matrix elements.
CKM matrix was set to identity.
All samples were generated using \texttt{MMTH2014nlo68cl} PDFs interfaced through \texttt{LHAPDF6}.

\subsection{Signal}
Signal events were generated using \texttt{MadGraph5\_aMC@NLO}~\textit{v2.4.2} and an  NLO capable UFO model.     
For the analysis sgluons masses in the range 0.9 - 1.5 TeV were considered. 
Renormalization and factorization scales were set equal to the sgluon mass.
Sgluons were then decayed into $t\bar t$ pairs (and further) using \texttt{MadSpin}~\cite{Artoisenet:2012st} generating all configurations that give two same-sign muons.
All spin correlations were preserved (at the LO).
Total branching ratio into these channels is given by 
$\text{BR}^2(W \to \mu \nu) (2-\text{BR}^2(W \to \mu \nu))$ where $\text{BR}(W \to \mu \nu) \approx 11$\%.
Partonic events were matched to parton shower using \texttt{MC@NLO}~\cite{Frixione:2002ik} prescription and \texttt{Pythia8}~\cite{Sjostrand:2014zea}~\textit{v219}.
\texttt{Pythia8} settings needed for consistent showering of \texttt{MC@NLO} events are described in Appendix~\ref{sec:pythia8_settings}.
Since there are no genuine NLO underlying event tunes in \texttt{Pythia8}, the default LO tune was used.

\subsection{Background validation}
          
Background samples were generated using \texttt{Sherpa}~\textit{v.2.2}~\cite{Gleisberg:2008ta}, with virtual matrix elements provided by \texttt{OpenLoops}~\textit{v1.3.1}~\cite{Cascioli:2011va} and evaluated using \texttt{CutTools}~\cite{Ossola:2006us,Ossola:2007ax} or \texttt{COLLIER}~\cite{Denner:2002ii,Denner:2005nn,Denner:2010tr,Denner:2016kdg}. 
$t\bar{t} \mu \nu_\mu$ (i.e. including $\mu^- \bar \nu_\mu$ and $\mu^+ \nu_\mu$ combinations) events were generated with up to 1 additional jet at NLO order and 3 jets at LO, while for $t\bar{t} \mu^+ \mu^-$ up to 1 and 2 jets, respectively, were generated.
Different multiplicities were merged using the MEPS@NLO technique \cite{Hoeche:2012yf,Gehrmann:2012yg}.
In case of $t\bar{t} \mu^+ \mu^-$ a generation cut on an invariant mass of the muon pair $m_{\mu^+ \mu^-} > 20$ GeV was applied.
Top quarks were then decayed in all ways that ensure two same-sign muons with spin correlations preserved at the LO as in the case of \texttt{MadSpin}.
The inclusive cross sections for those samples (including appropriate top-quarks decays) are 7.77 and 5.43 fb, respectively.
These predictions agree within (still very large) experimental uncertainties with the LHC measurements \cite{ATLAS-CONF-2016-003,CMS-PAS-TOP-16-009}.

The setup of \texttt{Sherpa} mostly follows standard settings. 
Here only the most important ones are mentioned. 
Samples were generated with \texttt{EXCLUSIVE\_CLUSTER\_MODE = 1} setting (meaning that only QCD splittings are considered when reconstructing parton shower history) to ensure that $t\bar{t} \mu \nu_\mu$/$t\bar{t} \mu^+ \mu^-$ is always identified as the core process.
Since ATLAS analysis uses jets with $p_T > 20$ GeV, the merging cut was set to 15 GeV.
Also, a default scale definition for the core process was used.

\section{Recasting current ATLAS 13 TeV analysis}

\begin{table}
  \centering
  \begin{tabular}{c||ccc|c}
    & SS muon pair & \# b-jets $\geq 3$ & $m_{\text{eff}} > 650$ GeV & $E_T^{\texttt{miss}} > 125$ GeV\\
    \hline
    \hline
    $t\bar{t} \mu \nu$ & 3.1876 & 0.0899 & 0.0198 & $0.0117 \pm 0.0006$\\
    $t\bar{t} \mu^+ \mu^-$ & 2.850 & 0.102 & 0.028 & $0.010 \pm 0.001$\\
    \hline
    $m_O = 0.90$ TeV & 1.352 & 0.707 & 0.629 & $0.424 \pm 0.002$\\
    $m_O = 1.00$ TeV & 0.6410 & 0.3324 & 0.3081 & $0.2172 \pm 0.0007$ \\
    $m_O = 1.25$ TeV & 0.1144 & 0.0569 & 0.0552 & $0.0426 \pm 0.0001$ \\
    $m_O = 1.50$ TeV & 0.02365 & 0.01109 & 0.01094 & $0.00897 \pm 0.00003$
  \end{tabular}
  \caption{Cut-flow analysis summary (numbers in fb).
      For brevity's sake, errors only for the final results are given.
      Errors are only statistical.
  \label{tab:cut-flow}}
\end{table}
\definecolor{Gray}{gray}{0.9}
\begin{table}
  \centering
  \begin{tabular}{c||c|c}
    & this analysis & ATLAS \\
    \hline
    \hline
    $t\bar{t} \mu \nu$ & $0.149 \pm 0.007$ & $0.10 \pm 0.05$\\
    $t\bar{t} \mu^+ \mu^-$ & $0.12 \pm 0.02$ & $0.14 \pm 0.06$\\
    \hline
    $m_O = 0.90$ TeV & $5.42 \pm 0.02$ & \cellcolor{Gray}\\
    $m_O = 1.00$ TeV & $2.781 \pm 0.009$ & \cellcolor{Gray}\\
    $m_O = 1.25$ TeV & $0.546 \pm 0.002$ & \cellcolor{Gray}\\
    $m_O = 1.50$ TeV & $0.1148 \pm 0.0003$ & \cellcolor{Gray} 
  \end{tabular}
  \caption{Final result of analysis (last column of \autoref{tab:cut-flow}) after multiplying by 3.2 fb$^{-1}$ of integrated luminosity and roughly a factor of 4 to account for all possible leptonic channels taken into account in the ATLAS analysis \cite{Aad:2016tuk} compared to column SRb3 of Tab.~5 of that analysis.
  \label{tab:my_cut-flow_vs_atlas}}
\end{table}

The ATLAS analysis of Ref.~\cite{Aad:2016tuk} targeted topologies with 2 same-sign leptons or 3 leptons, looking at 4 different signal regions.
In case of the production of sgluon-pair which then decays to top-quark pairs the interesting signal region is SR3b defined in Table 1 of \cite{Aad:2016tuk}.
To match experimental data as closely as possible, the detector response was parametrized using \texttt{Delphes} \cite{deFavereau:2013fsa} \textit{v3.3.2}. 

\begin{figure}
  \centering
  \includegraphics[width=0.6\textwidth]{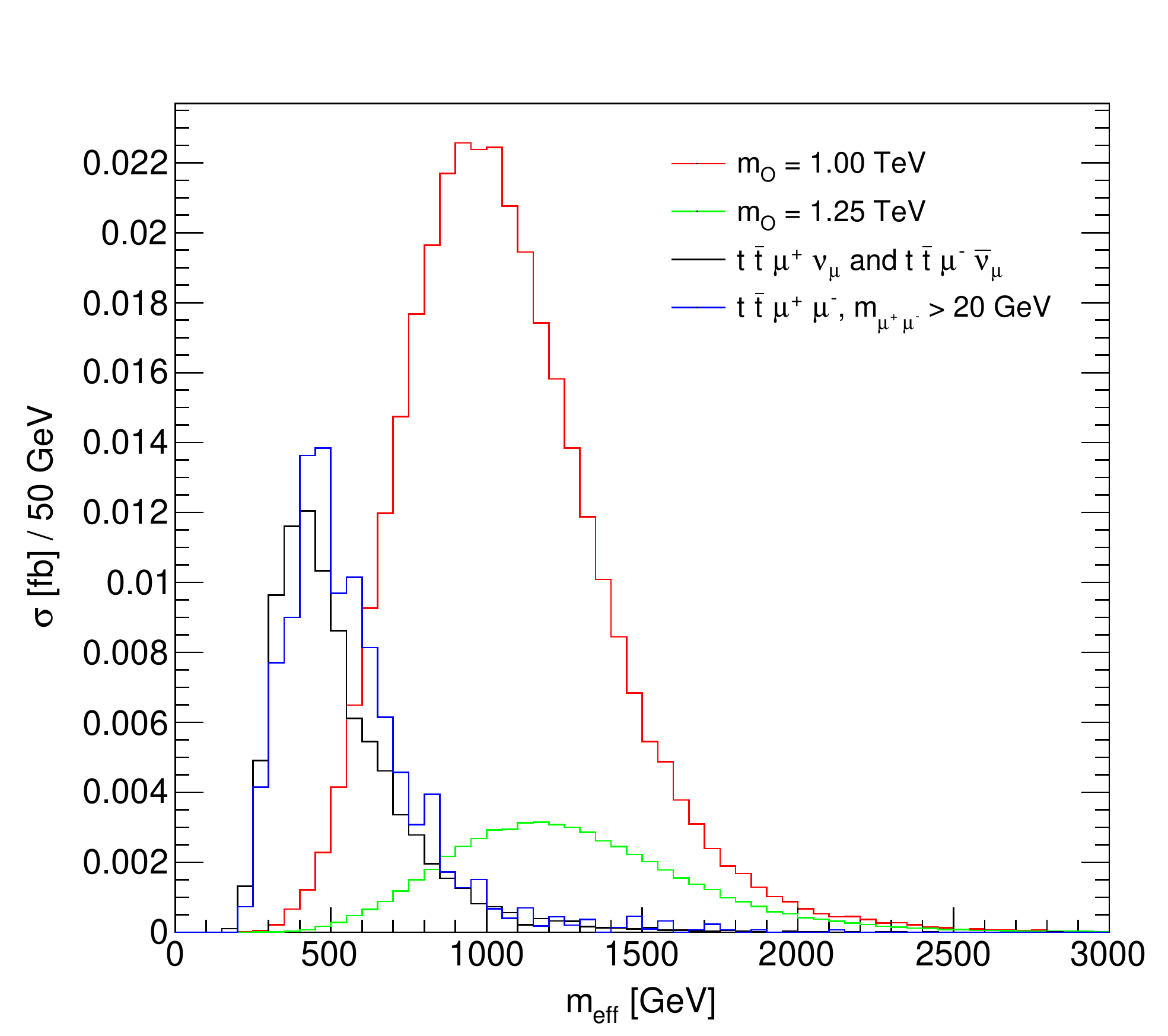}
  \caption{Effective mass spectrum after requiring 2 same-sign leptons and at least 3 $b$-tagged jets (see text for details) for the signal from 1 TeV sgluon pair and background from $t\bar t \mu^+ \nu_\mu$/$t\bar t \mu^- \bar \nu_\mu$ and $t\bar t \mu^+ \mu^-$.}
  \label{fig:eff_mass}
\end{figure}

\begin{figure}
  \centering
  \subfloat[]{\includegraphics[width=0.49\textwidth]{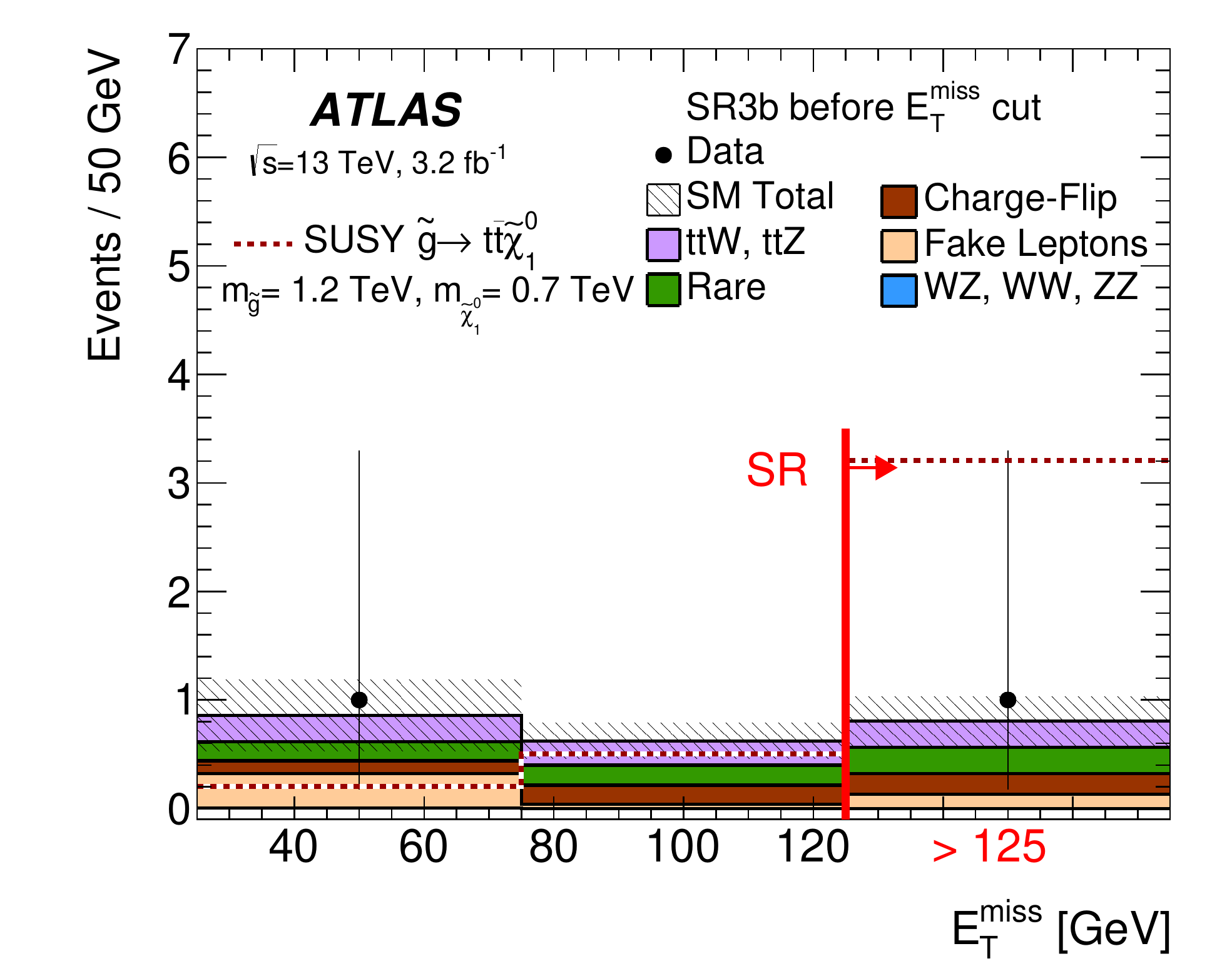}}
  \subfloat[]{
    \raisebox{1ex}{\includegraphics[width=0.49\textwidth]{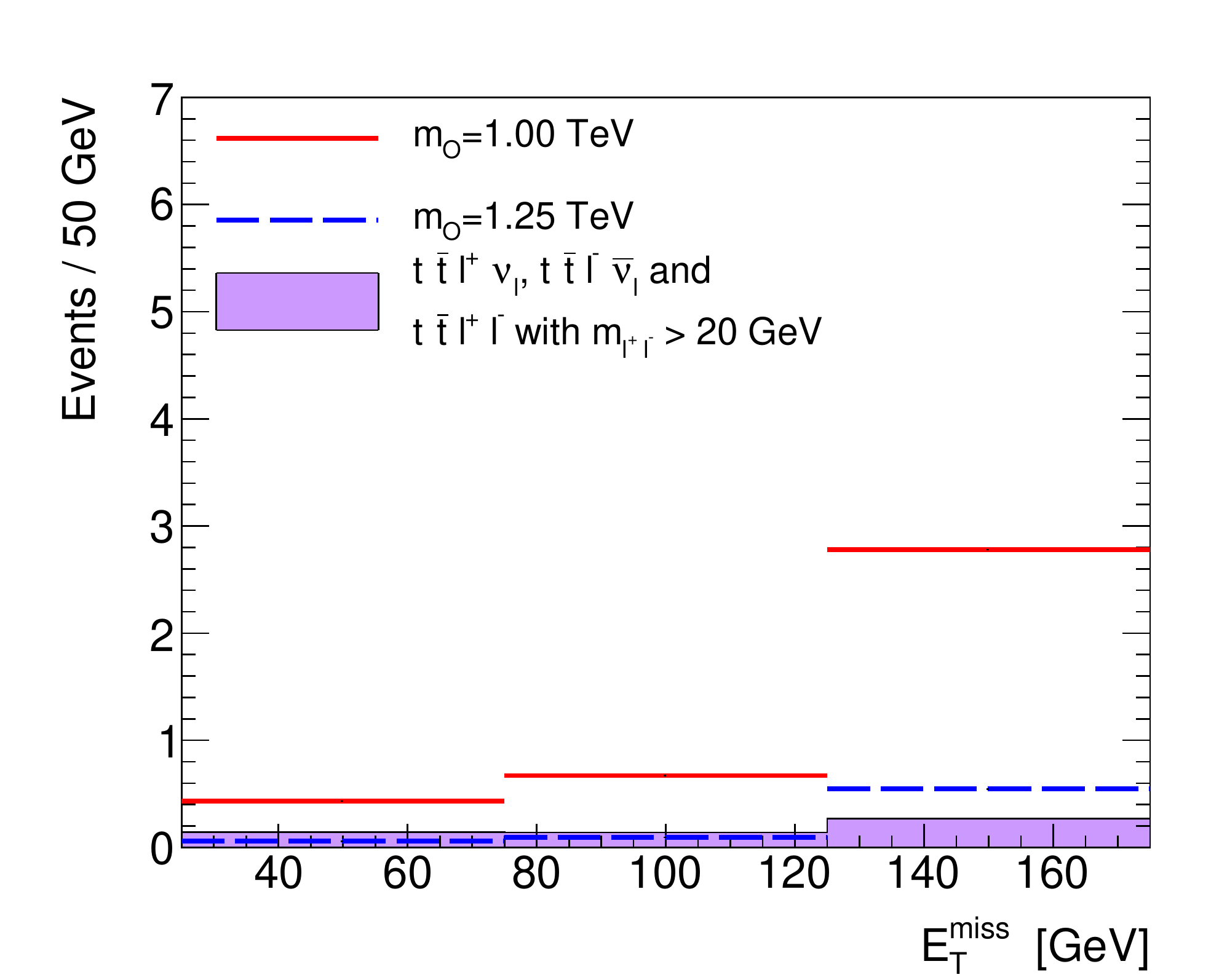}}
  }
  \caption{
  Spectrum of $E_T^{\text{miss}}$ by ATLAS~\cite{Aad:2016tuk} before applying the cut on it (a).
  Right panel (b) shows the analogous plot for $t\bar{t}W^\pm,t\bar{t}Z$ background based on the Monte Carlo simulation used in this work. 
  Red line in right panel b shows the superimposed signal from 1 TeV sgluon production. 
  \label{fig:atlas}}
\end{figure}

The following list gives a summary of \texttt{Delphes} detector card settings\footnote{The complete ATLAS detector card used in this analysis can be found in supplementary materials with the \texttt{arXiv} version of this work.} and applied cuts:
\begin{itemize}

\item[1] Muons are identified with the efficiency of 95\% if they have $p_T > 10$ GeV and $|\eta| < 1.5$ and 85\% if $1.5 < |\eta| < 2.7$.
Candidate muons are required to have $p_T > 20$ GeV and $|\eta| < 2.5$.
Candidate muons must also be isolated, that is have the scalar sum of the $p_T$ of tracks within a variable-size cone around the lepton, excluding its own track, less than 6\% of the muon $p_T$.
The isolation cone size is taken to be the smaller of 10 GeV/$p_T$ and 0.3 (where $p_T$ denotes the muon's transverse momentum).\footnote{\texttt{Delphes} Isolation module was modified to allow for a variable isolation cone size.}

\item[2] At least 3 b-tagged jets  reconstructed using anti-kt algorithm \cite{Cacciari:2008gp} from \texttt{FastJet}~\cite{Cacciari:2011ma,hep-ph/0512210} with $p_T > 20$ GeV and $|\eta|<2.5$ are required.
Jets are $b$-tagged if they are within $\Delta R_{jb} < 0.3$ of a $b$-quark which had $p_T^b > 5$ GeV and $|\eta_b| < 2.5$ with an efficiency \cite{ATL-PHYS-PUB-2015-022}
\begin{equation}
b\text{-tagging efficient} = \frac{24 \tanh(0.003 \cdot p_T)}{1+0.086 \cdot p_T} 
\end{equation}
Jet energy scale correction is applied according to the formula\footnote{JES is applied \textit{before} the requirement of $p_T > 20$ GeV.} 
\begin{equation}
 E_j \to  \sqrt{1 + (3 - 0.2 |\eta|)^2/p_T} \cdot E_j
\end{equation}
\item[3] Effective mass $m_{\text{eff}}$ of the event, defined as a scalar sum of $p_T$ of signal leptons, b-jets and missing $E_T$, must satisfy $m_{\text{eff}} > 650$ GeV.
The $m_{\text{eff}}$ spectra for the signal and $t \bar{t} \mu \nu$, $t \bar{t} \mu^+ \mu^-$ backgrounds are shown in Fig.~\ref{fig:eff_mass}.
\item[4] $E_T^{\texttt{miss}} > 125$ GeV
\end{itemize}
Table \ref{tab:cut-flow} shows the cross sections (in fb) for different processes passing this sequence of cuts (cuts are stacked, that is a cut in the $n$-th column also implies that cuts in $n-1$ first columns were applied).
Table \ref{tab:my_cut-flow_vs_atlas} then compares final numbers of background events, that is after multiplying last column of Tab.~\ref{tab:cut-flow} by 3.2~fb$^{-1}$ of integrated luminosity and roughly a factor of 4 to account for all possible leptonic channels taken into account in the ATLAS analysis, with the column SRb3 of Tab. 5 of Ref.~\cite{Aad:2016tuk}. 
The fact that the simplified analysis based on \texttt{Delphes} predicts roughly the same number of events for background coming from $t\bar{t} \mu \nu_\mu$/$t\bar{t} \mu^+ \mu^-$ production as ATLAS one is a check of its implementation.
Since a significant contribution to the background comes from elements which cannot be reliably simulated by Monte Carlo, like fake/non-prompt leptons and charge flips, the cuts used in the definition of SR3b could not be adapted.
To check the separating power of those cuts on the sgluon signal a plot after cuts on same-sign muon pair and number of $b$-jets was done.
Figure~\ref{fig:eff_mass} shows the spectrum of the effective mass for two sgluons masses: 1 and 1.25 TeV and backgrounds from $t\bar{t} \mu \nu_\mu$ and $t\bar{t} \mu^+ \mu^-$.
It is clear that cut of $m_{\text{eff}} > 650$ GeV used in the ATLAS analysis does also a good job in separating background from the sgluon signal.
For completeness I also show the numbers for background and signal events after effective mass cut but before the cut on missing $E_T$.
They are compared with original ATLAS plot in Fig.~\ref{fig:atlas} together with superimposed signal from 1 TeV sgluon.

\begin{figure}
  \centering
  \includegraphics{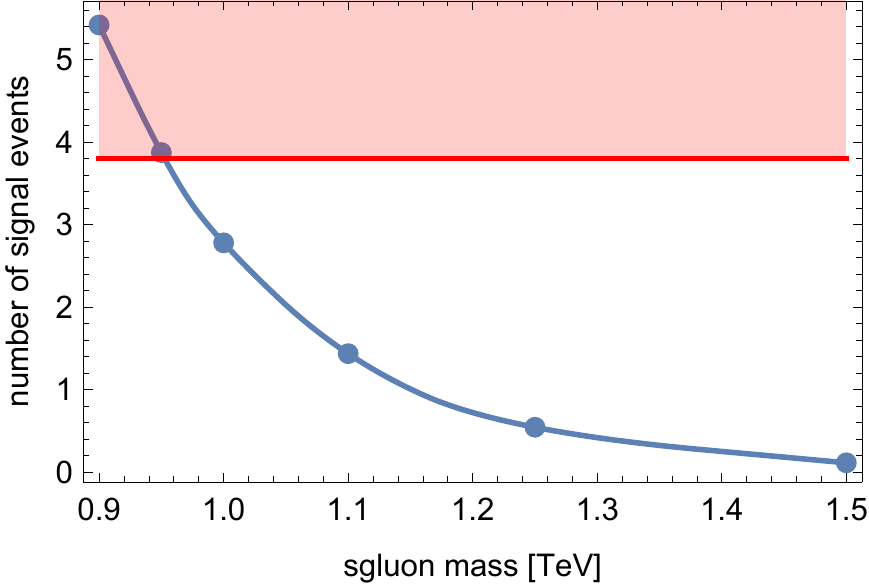}
  \caption{ 
    Predicted number of observed signal events as a function of the sgluon mass (blue points).
    Solid line shows interpolation between these points.
    Red region is excluded by ATLAS for SR3b at 95\% CL.
    Interpreted in the context of sgluon production it corresponds to a lower limit on the sgluon mass $m_O \lesssim 0.95$ TeV.
    \label{fig:mass_exclusion}
  }
\end{figure}

The 95\% CL observed upper limit on the number of signal (BSM) events in the SR3b is 3.8.
The predicted number of signal events for selected sgluon masses are given in Tab.~\ref{tab:my_cut-flow_vs_atlas}.
The ATLAS limit corresponds then to sgluons of mass in the range $0.9 < m_O < 1$ TeV.
To facilitate reading of its precise value, predicted numbers of signal events are plotted in Fig.~\ref{fig:mass_exclusion} together with the interpolation between them.
From this, sgluon masses < 0.95 GeV are excluded at 95\% CL.
This result is already on par with the 8 TeV ATLAS exclusion which was 1.06 TeV for the case of a complex sgluon (i.e. with cross section greater by a factor of 2).

The ATLAS experiment is supposed to gather 100 fb$^{-1}$ of integrated luminosity by the end of Run 2, roughly 30 times more than what is available currently.
Since statistical significance scales like a square-root of integrated luminosity, numbers in Tab.~\ref{tab:my_cut-flow_vs_atlas} suggest that even without further exploiting event kinematics and adapting cuts it should be possible to exclude (or discover) sgluons with masses up to $\lesssim 1.25$ TeV by the end of Run 2.

\section{Conclusions}

In this work I recast current ATLAS exclusion limits coming from the search of 4-top quark final state in events with same-sign leptons to the case of sgluon pair production.
Although sgluons decay to a top-quark pair without (typical in SUSY theories) presence of the invisible LSP assumed in the ATLAS analysis, cuts used turn out to work well also in this case.
Currently published data allow therefore to exclude sgluons with masses $\lesssim 0.95$ TeV, a result already on par with the 8 TeV exclusions.
Just from the increased statistics it should be therefore possible to push this limit up to 1.25 TeV by the end of Run 2.
Of course with an increased statistics experimental collaboration will be able to adapt the selection criteria to further exploit sgluon kinematics, pushing this exclusion even further.
We therefore encourage experimentalist to look into this.
\acknowledgments

I thank Frank Siegert for his help concerning \texttt{Sherpa}, Stefan Prestel for his constant support for \texttt{Pythia8}, Dominik Stöckinger and Philip Diessner for useful discussions about NLO (S)QCD corrections and Jan Kalinowski for suggesting the topic of this work. 
Work supported in part by the German DFG grant STO 876/4-1 and Polish National Science Centre under the decision UMO-2015/18/M/ST2/00518 (2016-2019).


\appendix
\section{\texttt{Pythia8} technical setup \label{sec:pythia8_settings} }

By default the final state shower algorithm in \texttt{Pythia8} is based on the dipole-style recoils. 
As stated in \texttt{Pythia8} manual, for \texttt{MC@NLO} where a full analytic knowledge of the shower radiation pattern in needed one has to switch to global recoil approach which does not contain color coherence phenomena (and hence factorizes).
A minimal set of settings needed to consistently shower \texttt{MC@NLO} events is then given by\footnote{See \texttt{Pythia8} manual at \url{http://home.thep.lu.se/~torbjorn/pythia82html/Welcome.html}, section \textit{Link to Other Programs $\to$ Matching and Merging $\to$  aMC@NLO Matching}. See also the discussion in Ref.~\cite{ATL-PHYS-PUB-2016-005}.}\\
\-\qquad \texttt{SpaceShower:pTmaxMatch = 1}\\
\-\qquad \texttt{SpaceShower:pTmaxFudge = 1.}\\
\-\qquad \texttt{SpaceShower:MEcorrections = off}\\
\-\qquad \texttt{TimeShower:pTmaxMatch = 1}\\
\-\qquad \texttt{TimeShower:pTmaxFudge = 1.}\\
\-\qquad \texttt{TimeShower:MEcorrections = off}\\
\-\qquad \texttt{TimeShower:globalRecoil = on}\\
\-\qquad \texttt{TimeShower:weightGluonToQuark = 1}\\
Those settings cannot be modified.
What can be chosen, though, is when to return from the global recoil mode to the dipole recoil.
Since color coherence phenomena are very important (see for example \cite{Chatrchyan:2013fha}), it is advantageous to switch back to dipole recoils already after the first emission. 
This can be done in two ways, setting \texttt{TimeShower:globalRecoilMode = 1} or \texttt{2}. 
Option 2 applies global recoil only if the first branching in evolution is a timelike splitting of a parton in an event with Born-like kinematics (the so called $\mathbb{S}$-events in the \texttt{MC@NLO} language), while for option 1 this is done both for Born-like ($\mathbb{S}$) and real-emission events ($\mathbb{H}$-events).
With option 2 the impact of global recoil should be minimal.
For options 1 and 2 a maximal number of splittings in the timelike shower with global recoil strategy should be set to 1 through \texttt{TimeShower:nMaxGlobalBranch} flag.
Also, to distinguish between $\mathbb{S}$ and $\mathbb{H}$ events, the number of color-charged particles for Born-like configurations must be given through \texttt{TimeShower:nPartonsInBorn} option.
The \texttt{MC@NLO} matching is done at the level of the hard process.
To that end, \texttt{Pythia8} removes decay chains generated by \texttt{MadSpin}
by traversing the event tree and identifying intermediate particles with status code \texttt{ISTUP}=$\pm 2$~\cite{Boos:2001cv} which have a single parent. 
\texttt{TimeShower:nPartonsInBorn}  then counts the number of remaining color-charged particles.
For the sgluon pair production I therefore set:
\\
\-\qquad \texttt{TimeShower:globalRecoilMode = 2}\\
\-\qquad \texttt{TimeShower:nMaxGlobalBranch = 1}\\
\-\qquad \texttt{TimeShower:nPartonsInBorn = 2}\\
\-\qquad \texttt{TimeShower:limitPTmaxGlobal = on}\\


\end{document}